\documentclass[aps,twocolumn,amsmath,superscriptaddress,amssymb,floatfix,preprintnumbers,nofootinbib,eqsecnum]{revtex4}

\usepackage{graphicx,subfigure,multirow}
\usepackage{longtable}
\usepackage{dcolumn}
\usepackage{bm}
\usepackage[all]{xy}
\usepackage{color}
\usepackage{amsmath}
\usepackage[usenames,dvipsnames]{xcolor}
\usepackage[unicode=true,pdfusetitle,
 bookmarks=true,bookmarksnumbered=false,bookmarksopen=false,citecolor=Turquoise,
 breaklinks=false,pdfborder={0 0 1},backref=false,colorlinks=true,pdfpagemode=FullScreen]
 {hyperref}

\usepackage{ulem}

\makeatletter

\newcommand{\fmslash}[2][0mu]{%
  \mathchoice
    {\fmsl@sh\displaystyle{#1}{#2}}%
    {\fmsl@sh\textstyle{#1}{#2}}%
    {\fmsl@sh\scriptstyle{#1}{#2}}%
    {\fmsl@sh\scriptscriptstyle{#1}{#2}}}
\newcommand{\fmsl@sh}[3]{%
  \m@th\ooalign{$\hfil#1\mkern#2/\hfil$\crcr$#1#3$}}
\makeatother

\newcommand{\lsim}{{\;\raise0.3ex\hbox{$<$\kern-0.75em\raise-1.1ex\hbox{$\sim$}}\;}}
\newcommand{\gsim}{{\;\raise0.3ex\hbox{$>$\kern-0.75em\raise-1.1ex\hbox{$\sim$}}\;}}

\newcommand{\beq}{\begin{equation}}
\newcommand{\eeq}{\end{equation}}
\newcommand{\bea}{\begin{eqnarray}}
\newcommand{\eea}{\end{eqnarray}}
\mathchardef\minus="002D

\usepackage{color}

\def\beq{\begin{equation}}
\def\eeq{\end{equation}}
\def\bea{\begin{eqnarray}}
\def\eea{\end{eqnarray}}
\def\eq#1{Eq.~(\ref{#1})}

\begin{document}
\title{Inelastic Boosted Dark Matter at Direct Detection Experiments}

\author{Gian F. Giudice}
\affiliation{Theoretical Physics Department, CERN, Geneva, Switzerland}
\author{Doojin Kim}
\affiliation{Theoretical Physics Department, CERN, Geneva, Switzerland}
\author{Jong-Chul Park}
\affiliation{Department of Physics, Chungnam National University, Daejeon 34134, Korea}
\author{Seodong Shin}
\affiliation{Enrico Fermi Institute, University of Chicago, Chicago, IL 60637, USA}
\affiliation{Department of Physics \& IPAP, Yonsei University, Seoul 03722, Korea}

\preprint{
\begin{minipage}[b]{1\linewidth}
\begin{flushright}
CERN-TH-2017-258 \\
EFI-17-24\\
 \end{flushright}
\end{minipage}
}

\begin{abstract}
We explore a novel class of multi-particle dark sectors, called Inelastic Boosted Dark Matter (iBDM). These models are constructed by combining properties of particles that scatter off matter by making transitions to heavier states (Inelastic Dark Matter) with properties of particles that are produced with a large Lorentz boost in annihilation processes in the galactic halo (Boosted Dark Matter). This combination leads to new signals that can be observed at ordinary direct detection experiments, but require unconventional searches for energetic recoil electrons in coincidence with displaced multi-track events. Related experimental strategies can also be used to probe MeV-range boosted dark matter via their interactions with electrons inside the target material.
\end{abstract}

\maketitle

\section{Introduction}
The existence of dark matter (DM), which is advocated by various cosmological and astrophysical observations, is mostly rooted in its gravitational interactions.
The strategy of direct detection experiments aims at revealing DM through its possible non-gravitational couplings to Standard Model (SM) particles, mostly by observing a recoiling of target material (henceforth called primary signature) which is induced by {\it elastic} scattering off of {\it non}-relativistic DM.
A variation in this search scheme is to look for {\it inelastic} scattering signals, imagining that a DM particle scatters off to an excited state (if kinematically allowed) along with a target recoiling whose spectrum differs from that in the elastic scattering~\cite{TuckerSmith:2001hy}.
One may instead focus on secondary visible particles which are potentially involved in de-excitation of the excited state, e.g., X-ray photon
in neutrinoless double beta decay experiments~\cite{Pospelov:2013nea}. However, it is generally assumed that observing {\it both} primary and secondary signatures is unlikely due to inadequate DM kinetic energy to overcome the relevant thresholds simultaneously.
Many DM direct detection experiments have searched for DM signals, but null observation merely sets stringent bounds on DM models.

These negative results had a profound impact on the field. Traditional experimental searches were designed to target the range of DM masses around 100~GeV and weak-scale scattering cross sections off nuclei, which correspond to the parameter region preferred by Weakly Interacting Massive Particle (WIMP) models. However, these experimental techniques rapidly lose sensitivity below masses of a few GeV and this has ignited a great interest in the exploration of the sub-GeV DM domain. This exploration in the low-mass region is not a simple extrapolation of traditional searches. On the contrary, sub-GeV DM experimental searches are characterized by completely new techniques~\cite{Battaglieri:2017aum}. The theoretical approach has changed significantly. Instead of building realistic models addressing weak-scale physics and obtaining the DM as a byproduct, the focus is now in inventing new experimental strategies to hunt for DM in unexplored windows of parameter space and propose new signatures, quite distinct from those traditionally looked for. The emphasis today is on the experimental searches, rather than the theoretical model building. The models for sub-GeV DM may superficially appear less ambitious and more {\it ad hoc} than traditional WIMP models, but their primary role is to motivate interesting and unconventional experimental searches. Moreover, the WIMP assumption of a single DM particle -- especially when compared with the complexity of ordinary matter -- may be an oversimplification, and the exploration of more complex dark sectors is today well justified.
In this paper, we will follow this perspective and propose novel signatures from multi-particle dark sectors that can be looked for at DM direct detection experiments, but that are completely distinct from the traditional expectation.

The paper is organized as follows. In Sec.~\ref{sec:first} we discuss the general strategy while in Sec.~\ref{sec:model} we briefly describe the benchmark DM model adopted throughout this paper. We then study key kinematic features in signal events, including energy spectra and decay lengths of visible particles in Sec.~\ref{sec:kinematics}. Section~\ref{sec:results} contains our main results including detection strategy, background consideration, and phenomenology. We summarize our results in Sec.~\ref{sec:outlook}. Finally, the appendices are reserved for benchmark DM model details and useful formulae for the scattering and decay processes.

\section{General strategy} \label{sec:first}

An alternative approach to traditional searches for DM is offered by the so-called {\it boosted dark matter}~\cite{Agashe:2014yua}.
The underlying DM models hypothesize a dark sector comprising of two DM species
with a hierarchical mass spectrum: the heavier and the lighter species are denoted by $\chi_0$ and $\chi_1$, respectively.\footnote{The stability of each of them is ensured by two separate symmetries such as $Z_2 \times Z_2'$ or U(1)$'\times$ U(1)$''$. See, for example, the model in Ref.~\cite{Belanger:2011ww}.}
Typical scenarios assume that $\chi_0$ has no direct coupling to SM particles but pair-annihilates into two $\chi_1$'s which directly communicate with SM particles.
Their respective relic abundances are determined by the ``assisted'' freeze-out mechanism~\cite{Belanger:2011ww} rendering the heavier (lighter) a dominant (sub-dominant) DM component.
In the present universe, the {\it boosted} $\chi_1$ can be produced via pair-annihilation of $\chi_0$ in the galactic halo, leading to a total flux~\cite{Agashe:2014yua}
\bea
\mathcal{F} &=& 1.6 \times 10^{-4}\, {\rm cm}^{-2}{\rm s}^{-1} \nonumber \\ &&
\left( \frac{ \langle \sigma v \rangle_{0 \to 1}}{5\times 10^{-26}\, {\rm cm}^3{\rm s}^{-1}}\right) \left( \frac{\rm GeV}{m_{0}}\right)^2 \, ,
\label{eq:flux}
\eea
where the reference value for $\langle \sigma v \rangle_{0 \to 1}$, the velocity-averaged annihilation cross section of $\chi_0 \chi_0 \to \chi_1 \chi_1$, corresponds to a correct DM thermal relic density for $\chi_0$.

For $\chi_0$ of weak-scale mass (i.e., $\sim 100$ GeV),
the incoming flux of lighter DM $\chi_1$ (near the earth) is as small as $\mathcal O(10^{-8} \,{\rm cm}^{-2} {\rm s}^{-1})$
so that large volume neutrino detectors such as Super-Kamiokande, Hyper-Kamiokande, and Deep Underground Neutrino Experiment are preferred in search for elastic signatures~\cite{Agashe:2014yua,Berger:2014sqa,Kong:2014mia,Alhazmi:2016qcs} or inelastic signatures~\cite{Kim:2016zjx}. Very recently, the Super-Kamiokande Collaboration has reported the first results in the search for high-energy electrons ($\geq 0.1$ GeV) induced by elastic scattering of boosted DM $\chi_1$~\cite{Kachulis:2017nci}. Related searches can be conducted at fixed target experiments, with active production of boosted dark matter~\cite{Kim:2016zjx,deNiverville:2011it,Izaguirre:2014dua,Izaguirre:2017bqb}.

On the other hand, if the dominant relic component $\chi_0$ takes a mass in the sub-GeV regime, the large-volume neutrino detectors mentioned above may not be ideal to look for the signatures by $\chi_1$ due to their relatively high threshold energies (e.g., several tens to a hundred MeV).
Moreover, noting from Eq.~\eqref{eq:flux} that the $\chi_1$ flux is inversely proportional to the mass square of $\chi_0$, we observe that it can increase by 4--6 orders of magnitude with sub-GeV/GeV-range $\chi_0$ DM, while the resulting relic density is still in agreement with the current measurement.
Hence, it is quite natural to pay attention to relatively small (fiducial) volume detectors {\it but} with a low threshold energy: for example, conventional DM direct detection experiments.
We will show that current and future DM direct detection experiments such as XENON1T~\cite{Aprile:2017iyp, Aprile:2017aty}, DEAP-3600~\cite{Boulay:2012hq, Amaudruz:2014nsa, Amaudruz:2017ekt}, and LUX-ZEPLIN (LZ)~\cite{Mount:2017qzi} may possess sufficient sensitivity to signals caused by boosted (lighter) DM of MeV-range mass.
A beginning effort was made in Ref.~\cite{Cherry:2015oca}; the authors assumed coherent scattering of nuclei by the boosted DM which arises  
in leptophobic scenarios (e.g., gauged baryon number or Higgs portal models), and reinterpreted the results from even smaller-volume detectors like LUX.

However, as we will discuss later, 
we observe that the MeV-range $\chi_1$-electron scattering can compete with the corresponding interaction with nucleons, unless the coupling associated with the electron is suppressed in specific model frameworks.
Therefore, a dedicated study on boosted light $\chi_1$-induced signatures involving the electron recoil (ER) is also required in probing boosted DM scenarios at conventional direct detection experiments, in which ordinary ER (lying in the keV to sub-MeV regime) is usually rejected due to a large rate of expected backgrounds.
However, we claim that the signal induced by boosted DM can be quite energetic (above a few to tens of MeV) and usually leaving an appreciable track in the detector system, which is clearly distinctive from conventional background events associated with ER.
In this sense, DM direct detection experiments should be highlighted as {\it discovery machines} of light boosted DM.

\begin{figure}[tbp]
\centering
\includegraphics[width=8.4cm]{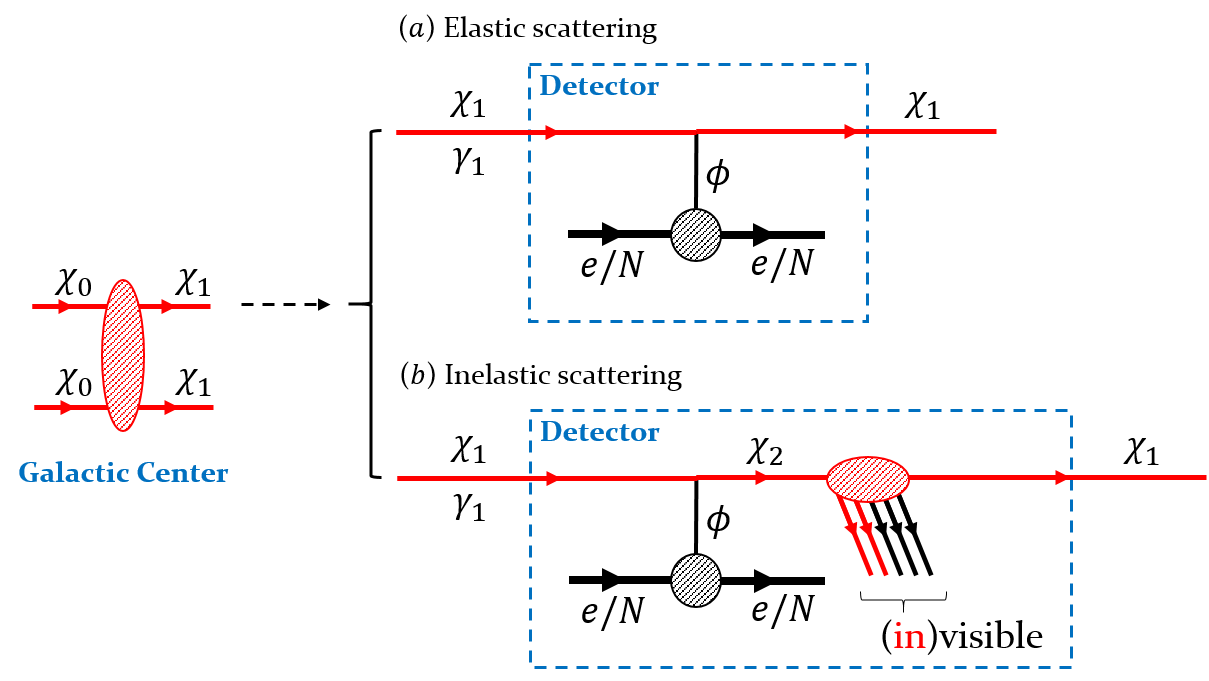}
\caption{\label{fig:scenario} The ordinary boosted DM (upper part) and iBDM (lower part) scenarios with the relevant DM-signal processes under consideration. }
\end{figure}

In this paper, we explore Inelastic Boosted Dark Matter (iBDM) as a novel paradigm for DM and discuss the characteristic signals that can be produced at direct detection experiments. We
take a comprehensive approach, considering both elastic and inelastic scatterings of {\it boosted} DM which are displayed in the upper and the lower panels of Fig.~\ref{fig:scenario}, correspondingly.
In particular, the latter possibility involves the process in which an incident DM particle $\chi_1$, produced by pair-annihilation of heavier DM $\chi_0$ (at say, the Galactic Center) with a boost factor $\gamma_1=m_{0}/m_{1}$, scatters off to a {\it heavier, unstable} dark sector particle $\chi_2$ together with a target recoil ($e$ or $p$) via a mediator $\phi$ exchange. This is what we refer to as primary process.
The $\chi_2$ then disintegrates back into $\chi_1$ and some other decay products which may include SM particles. This is called the secondary process.
If (at least) part of the secondary signature is observable in the same detector complex (blue dashed boxes in Fig.~\ref{fig:scenario}), the correlation between the primary and the secondary processes can be an additional, powerful handle to identify DM events from backgrounds
as well as to distinguish the iBDM scenario from ordinary boosted DM (elastic channel).
Furthermore, the secondary signal can be substantially displaced from the primary vertex over the position resolutions of the detector, depending on the parameter choice; this can be considered as unambiguous evidence for an inelastic scattering process.
From all the considerations above, we expect that good sensitivity to DM signals with ER can be achieved at current and future DM direct detection experiments.

\section{Benchmark dark matter models} \label{sec:model}
While the framework illustrated in Fig.~\ref{fig:scenario} is quite generic, we employ a concrete benchmark DM model throughout this paper in the spirit of the simplified model approach. The dark sector of our model, beyond the particle $\chi_0$ constituting the dominant DM component, contains at least two fermionic states $\chi_1$ and $\chi_2$ with masses $m_{1,2}$ (such that $m_{2}>m_{1}$). The communication between the SM and dark sector occurs through a (massive) dark photon $X$, playing the role of the generic mediator $\phi$. The relevant terms in the Lagrangian include the following operators:
\bea
\mathcal{L} \supset &-&\frac{\epsilon}{2}F_{\mu\nu}X^{\mu\nu}\nonumber \\
&+&g_{11}\bar{\chi}_1\gamma^{\mu}\chi_1 X_\mu +\left( g_{12}\bar{\chi}_2\gamma^{\mu}\chi_1 X_\mu + {\rm h.c.}
\right). \, \label{eq:lagrangian}
\eea
The first term describes the kinetic mixing between U(1)$_{\textrm{EM}}$ and U(1)$_X$~\cite{Okun:1982xi,Galison:1983pa,Holdom:1985ag,Huh:2007zw,Pospelov:2007mp,Chun:2010ve,Park:2012xq,Belanger:2013tla} parameterized by the small number $\epsilon$, with $F_{\mu\nu}$ and $X_{\mu\nu}$ being the field strength tensors for the ordinary and dark photon, respectively.\footnote{One can also think about a model where $\chi_1$ and $\chi_2$ transform under different representations of a dark gauge group with a mass mixing term,
inspired by the ideas in~\cite{Kim:2010gx,Dermisek:2015oja}.}
The couplings $g_{11}$ and $g_{12}$ in Eq.~\eqref{eq:lagrangian} measure the strength of the diagonal and off-diagonal currents, respectively.
The former is responsible for the elastic processes, usually considered in the phenomenology of boosted DM scenarios~\cite{Agashe:2014yua,Berger:2014sqa,Kong:2014mia,Alhazmi:2016qcs,Cherry:2015oca}.
The latter leads to the inelastic processes studied in this paper.  
The relative size of $g_{11}$ and $g_{12}$ is a completely model-dependent issue. As summarized in Appendix A, it is easy to conceive models in which diagonal interactions are highly suppressed or even vanishing, irrespectively of the mass splitting between $\chi_1$ and $\chi_2$, both for fermion and scalar particles. Therefore, there are scenarios in which inelastic processes are the leading experimental signature.

Our considerations can be straightforwardly generalized to other models. The dark sector could contain scalar particles, instead of fermions. The mediator $\phi$ between the SM and dark sectors, instead of the gauge boson considered here, could be a scalar (e.g., as in Higgs portal models~\cite{McDonald:1993ex,Kim:2008pp,Kim:2009ke}). Another possibility is to use a SM particle as mediator. This is the case for two chiral fermions $\chi_{1,2}$ with an effective magnetic dipole interaction. The ordinary photon acts as mediator and the chiral structure of the magnetic interaction insures the presence of only off-diagonal couplings, leading to the interesting conclusion that only inelastic processes are allowed.

We do not elaborate on details regarding possible cosmological constraints on the various models from nucleosynthesis or cosmic microwave background measurements, as they are beyond the scope of this paper.
Nevertheless, we have checked that the parameter space of interest for our benchmark model in Eq.~\eqref{eq:lagrangian} is cosmologically safe.

\section{Kinematic features} \label{sec:kinematics}

In this section, we discuss the key kinematic features in signal events. Although we will focus on the inelastic scattering, some of the relevant features for the elastic scattering can be readily obtained in the limit $m_{2} = m_{1}$.

\begin{figure}[tbp]
\centering
\includegraphics[width=4.2cm]{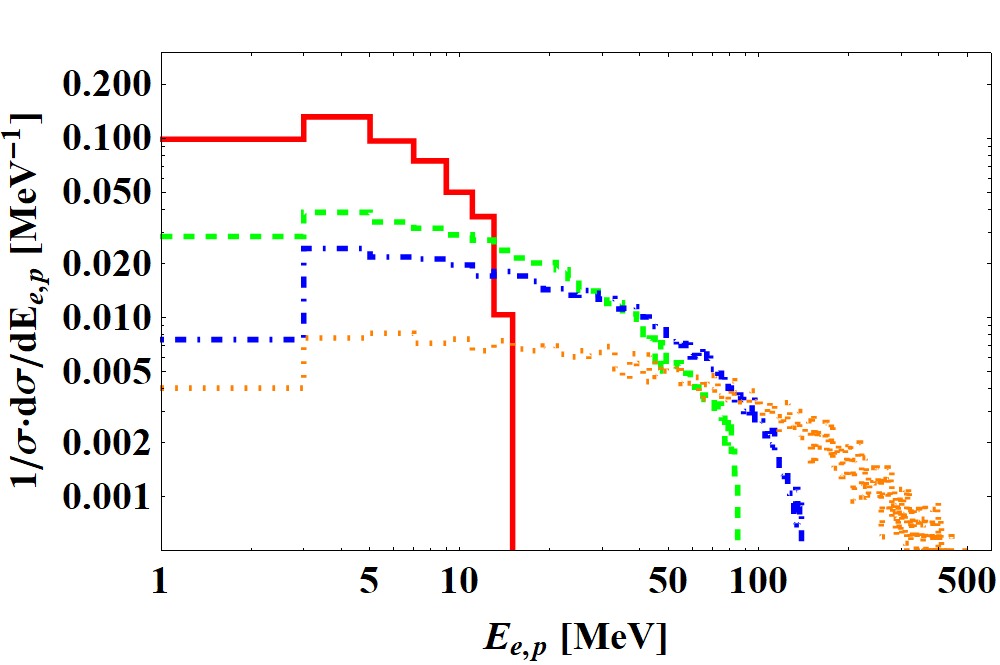}
\includegraphics[width=4.2cm]{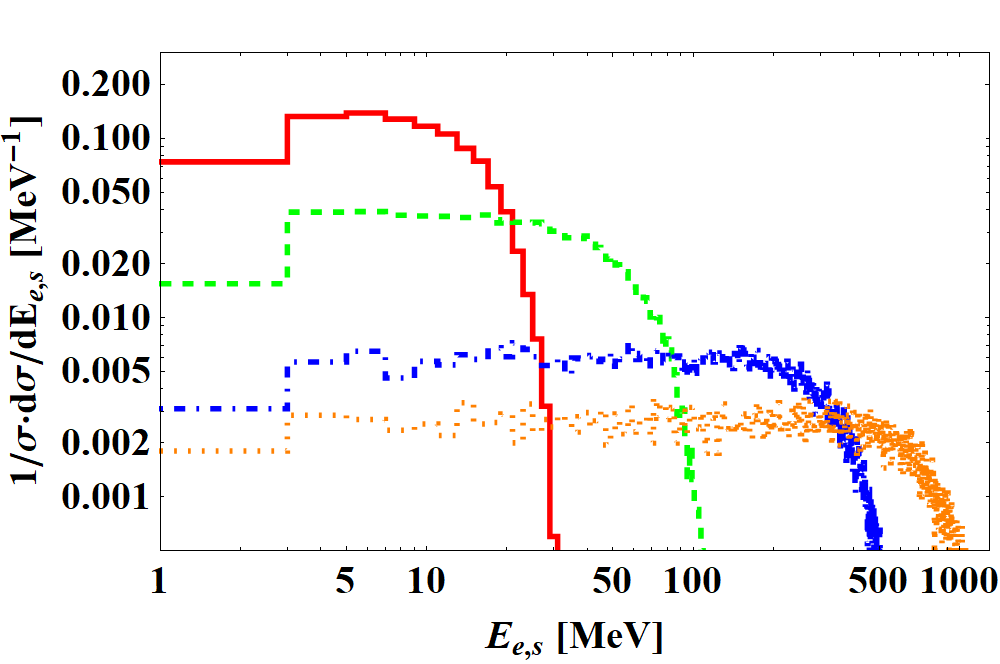}
\begin{tabular}{l | c c c c c}
 & \,\, $m_1$ \,\,& \,\,$m_2$\,\, & \,\,$m_X$\,\, & \,\,\,\,$\gamma_1$\,\,\,\, & $\epsilon$ \\
\hline \hline
ref1 (red solid) & 2 & 5.5 & 5 & 20 & $4.5\times 10^{-5}$\\
ref2 (green dashed) & 3 & 8.5 & 7 & 50 & $6\times 10^{-5}$ \\
\hline
ref3 (blue dot-dashed) & 20 & 35 & 11 & 50 & $7\times 10^{-4}$ \\
ref4 (orange dotted) & 20 & 40 & 15 & 100 & $6\times 10^{-4}$
\end{tabular}
\caption{\label{fig:spectrum} Predicted energy spectra of the primary (upper-left panel) and secondary (upper-right panel) $e^-$ and/or $e^+$ for four reference points, whose parameters are tabulated in the lower panel.
Here $g_{12}$ is set to unity and all mass quantities are in MeV.}
\end{figure}

We begin by considering the mass range of interest. In order to obtain a large flux of incoming lighter DM particles ($\chi_1$), we choose masses of the heavier DM ($\chi_0$) in the sub-GeV/GeV-range, see Eq.~\eqref{eq:flux}.
To obtain a sufficiently large boost factor in $\chi_1$ production, we choose the MeV range for $\chi_1$ and $\chi_2$.
If $\delta m\equiv m_{2} - m_{1}$ is smaller (larger) than $m_X$, the $\chi_2$ decay happens via an off-shell (on-shell) dark photon, which is denoted as off-shell (on-shell) scenario.

Note that there is a maximum allowed value of $\delta m$ for a given combination of incoming $\chi_1$ with energy $E_{1}(=\gamma_1 m_{1})$ and target mass $m_e$ 
\bea
\delta m &\le& \sqrt{ (m_1 + m_e)^2 + 2 (\gamma_1 -1) m_1 m_e}\nonumber \\ &&- (m_1 + m_e)\, . \label{eq:maxreach}
\eea
This implies that the parameter space for the on-shell dark photon scenario is rather limited. The condition $\delta m > m_X$ is more easily satisfied for large boost factor (i.e. for large $E_1$), which requires heavier $\chi_0$ (say, at the weak scale) and thus a small flux of $\chi_1$. 

For both the on-shell and off-shell scenarios,
 one obtains a (primary) recoil electron and an electron-positron pair in the final state. Here we focus on a three-electron final state, but our study can be extended to other signatures. One interesting example is the electron-photon final state from $\chi_2 \rightarrow \chi_1 \gamma$ by a magnetic dipole operator~\cite{Chang:2010en}.

The differential recoil spectrum for the process under consideration was thoroughly investigated in Ref.~\cite{Kim:2016zjx}, and we simply quote some key formulae in Appendix B for reference. We focus on 
the $\chi_1$ scattering cross section with electrons, but scattering with protons can be recovered by replacing
 $m_e$ with $m_p$ (and including the appropriate form factors).
Let us consider the limit of $m_e \ll m_X \ll m_p$.
In this limit, the propagator of the mediator $X$ suppresses the scattering cross section off electrons (see Eq.~\eqref{eq:matrixX} in Appendix B) proportionally to $(m_X^2 + 2 m_e E_e)^{-2}$, which implies that the lighter $X$ is, the larger the cross section.
On the other hand, the $X$ propagator gives a suppression of the scattering cross section off protons  proportional to $m_p^{-2} (m_p - E_p)^{-2}$: thus, the rate is smaller as the momentum transfer grows.
This situation is typical when the energy of the incoming light DM ($E_1$) is large but the energy of the outgoing light DM $\chi_1$ (for the elastic scattering case) or of the heavier dark-sector particle $\chi_2$ (for the inelastic scattering case) is small.
With this simple argument, we see that scattering off electrons can compete with 
scattering off protons for $m_e \ll m_X \ll m_p$ and for large enough boost factor $\gamma_1$. We have explicitly checked that this is indeed the case in the majority of the relevant parameter space, including our reference points presented shortly. 

In the upper-left panel of Fig.~\ref{fig:spectrum} we show the (unit-normalized) electron-recoil energy spectra for four reference points. The parameter choice of these reference points (which satisfy the various phenomenological and cosmological constraints) is described in the lower panel of Fig.~\ref{fig:spectrum}. The histograms are generated using the code \texttt{MG5@aMC}~\cite{Alwall:2014hca}.
We see that the typical electron-recoil kinetic energy is above $\mathcal{O}$(1 MeV), which is much greater than 
the usual threshold of direct detection experiments, $\mathcal{O}$(1 keV).
Nevertheless, if $\delta m$ is sizable, a large portion of the energy carried by $\chi_1$ is transferred to the mass and momentum of $\chi_2$, and thus $\chi_2$ usually comes out with a sizable boost factor $\gamma_2$.

\begin{figure}[tbp]
\centering
\includegraphics[width=5.2cm]{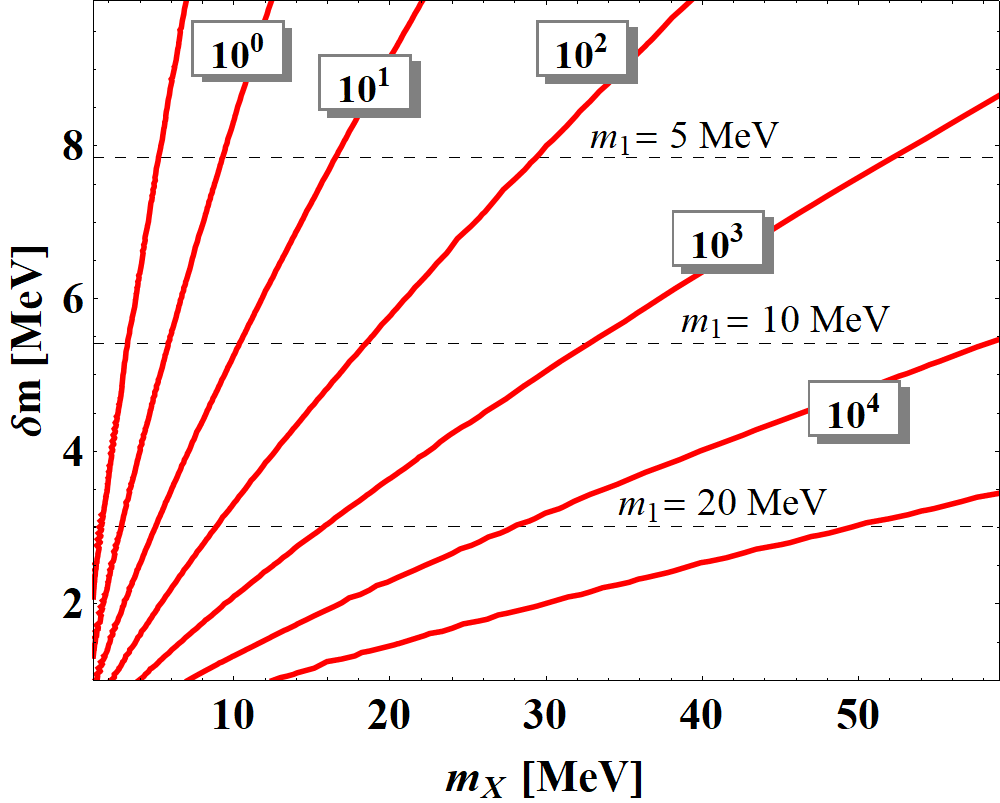}
\caption{\label{fig:mXdelm} Laboratory-frame decay length of $\chi_2$ (in units of cm) in the $m_X$--$\delta m$ plane for $g_{12}=1$, $\epsilon = 10^{-3}$, $\gamma_2=10$.
The dashed lines represent the maximum allowed $\delta m$ for given $m_1$ and $E_1=150$ MeV. }
\end{figure}

Let us turn to the secondary process. For a three-body process, the $\chi_2$ decay length can be appreciable and the associated decay width $\Gamma_2$ is
\bea
\Gamma_2 \approx \frac{ \epsilon^2 \alpha\, g_{12}^2	}{15\pi^2m_X^4}(\delta m)^5\,, \label{eq:app}
\eea
where $\alpha$ is the electromagnetic fine structure constant and we assumed $m_e \ll \delta m \ll m_2 \ll m_X$  (for the exact expression see Appendix B).
Note that the approximate expression in Eq.~\eqref{eq:app} is valid for both scalar and fermion DM scenarios.
The laboratory-frame mean decay length of $\chi_2$ (denoted as $\ell_{2,\textrm{lab}}$) is
\bea
\ell_{2,\textrm{lab}}=\frac{c \gamma_2}{\Gamma_2} &\sim& 16\hbox{ cm} \left(\frac{10^{-3}}{\epsilon} \right)^2  \left( \frac{1}{g_{12}}\right)^2  \nonumber \\
&\times& \left( \frac{m_X}{30 \hbox{ MeV}} \right)^4  \left( \frac{10 \hbox{ MeV}}{\delta m} \right)^5  \frac{\gamma_2}{10}\,.~~ \label{eq:decaylength}
\eea
This expression shows that there is an interesting range of parameters in which the physical separation between primary and secondary processes is sizable, but the secondary decay process still takes place within the detector (fiducial) volume. Contours of $\ell_{2,\textrm{lab}}$ are shown in
Fig.~\ref{fig:mXdelm}, in unit of cm, in the $m_X$--$\delta m$ plane for $g_{12}=1$, $\epsilon = 10^{-3}$, $\gamma_2 =10$.
For other choices, one can easily estimate the $\chi_2$ decay length using the scaling shown in Eq.~\eqref{eq:decaylength}. The dashed lines in Fig.~\ref{fig:mXdelm} show the maximum value of $\delta m$ determined by Eq.~\eqref{eq:maxreach}, for three values of $m_1$ and $E_1=150$ MeV.

In the case of two-body $\chi_2$ decay into an on-shell dark photon, the displacement of the secondary process may be caused by the subsequent $X$ decay.
The decay width $\Gamma_X$ is given by
\bea
\Gamma_X = \frac{\epsilon^2 \alpha\, m_X}{3}\left(1+\frac{m_e^2}{m_X^2}\right) \sqrt{1-\frac{4m_e^2}{m_X^2}}\,,
\eea
which corresponds to the following mean decay length
\bea
\ell_{X,\textrm{lab}}=\frac{c \gamma_X}{\Gamma_X} \sim 0.4 \hbox{ cm} \left(\frac{10^{-4}}{\epsilon} \right)^2 \left( \frac{20 \textrm{ MeV}}{m_X}\right)  \frac{\gamma_X}{10}\,.
\eea
This shows that the relevant displaced vertex is generally less appreciable than that in the previous case, since the position resolutions in XENON1T and LZ are
$\mathcal O (1\,{\rm cm})$~\cite{Aprile:2017iyp,Aprile:2017aty,Mount:2017qzi}.\footnote{Strictly speaking, at XENON1T, the resolution in the $z$-direction is about $\sim 0.1$ mm while in the radial direction is $\sim 2$ cm.}
The position resolution of DEAP-3600 is in general worse since only the prompt scintillation signal (S1) is measured.
However, its expected position resolution can be better than 6.5 cm with a sophisticated maximum likelihood fitter~\cite{Boulay:2012hq}.
Nevertheless, even for measurable $\ell_{X,\textrm{lab}}$, pinning down the underlying dynamics (i.e., on-shell vs. off-shell scenarios) is rather non-trivial.

The (unit-normalized) energy spectra of secondary decay products $e^{\pm}$ are shown in the upper-right panel of Fig.~\ref{fig:spectrum},  for our reference points. Not surprisingly, their typical energy is greater than the threshold of conventional DM direct detection experiments.
Since the energies of both the recoiling electron and the electron-positron pair from the secondary vertex can easily be at the level of $\mathcal O ({\rm MeV}$--$100$  MeV), ER-like signals are expected to produce between 1 and 3 energetic $e^\pm$ in the final state.

\section{Results} \label{sec:results}

\subsection{Benchmark detectors and detection strategy}
\begin{table}[tbp]
\centering
\begin{tabular}{c | c c c c }
Experiment & Geometry & $(r,~h)$ or $r$ [cm]& Mass [t] & Target \\
\hline \hline
XENON1T & Cylinder &(38,~76) & 1.0  & LXe \\
DEAP-3600 & Sphere & 72 & 2.2 & LAr \\
\hline
LZ & Cylinder &(69,~130) & 5.6 & LXe
\end{tabular}
\caption{\label{tab:exp} Summary of benchmark detector specifications for XENON1T~\cite{Aprile:2017aty}, DEAP-3600~\cite{Amaudruz:2014nsa}, and LZ~\cite{Mount:2017qzi}. The size corresponds to the fiducial volumes.
}
\end{table}

As mentioned in the Introduction, XENON1T~\cite{Aprile:2017iyp, Aprile:2017aty}, DEAP-3600~\cite{Boulay:2012hq, Amaudruz:2014nsa, Amaudruz:2017ekt}, and LZ~\cite{Mount:2017qzi} are employed as our benchmark experiments due to their large target masses, the first two of which are in operation while the last is projected in the near future.
We summarize their key characteristics in Table~\ref{tab:exp}, where the size corresponds to the fiducial volumes. The XENON1T and LZ experiments utilize a dual-phase time projection chamber filled with a liquid xenon (LXe) target where scattering events create both a prompt scintillation signal (called S1) and free electrons which further move to the gas phase Xe above the liquid Xe by a drifting electric field to generate a scintillation signal (called S2). The photomultiplier tubes (PMTs) in the top and bottom plates sense the scintillation photons.
The time difference between S1 and S2 is used to measure the depth (say, $z$ coordinate) of a scattering point by $\sim 0.1$ mm resolution~\cite{com}, while a likelihood analysis of S2 allows for the identification of the associated $xy$ position with 1--2 cm resolution~\cite{Aprile:2017iyp,Aprile:2017aty,Mount:2017qzi}.\footnote{A more dedicated likelihood analysis can attain $<1$ cm resolution in the $xy$ plane for higher energy sources~\cite{Akerib:2017riv}.}
On the other hand, DEAP-3600 uses a single-phase liquid argon (LAr) detector which records only S1.
We hereafter focus on the detectors at XENON1T and LZ unless specified otherwise. However, since both LXe and LAr detectors are similar to each other, we will also comment on difference/similarity between them wherever needed.

We recall that our DM signals are quite energetic. To the best of our knowledge, no dedicated detector studies of high-energy signals have been performed. So, the results shown in this paper should be regarded only as theoretical assessments meant to encourage detailed experimental analyses.

\begin{figure}[tbp]
\centering
\includegraphics[width=4.2cm]{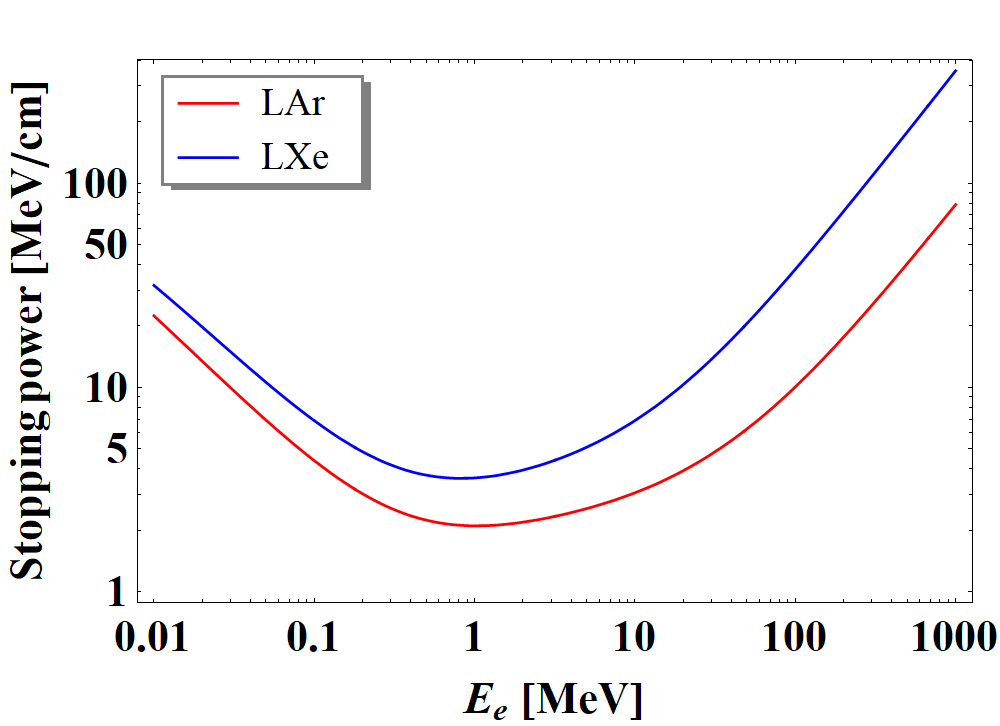}
\includegraphics[width=4.2cm]{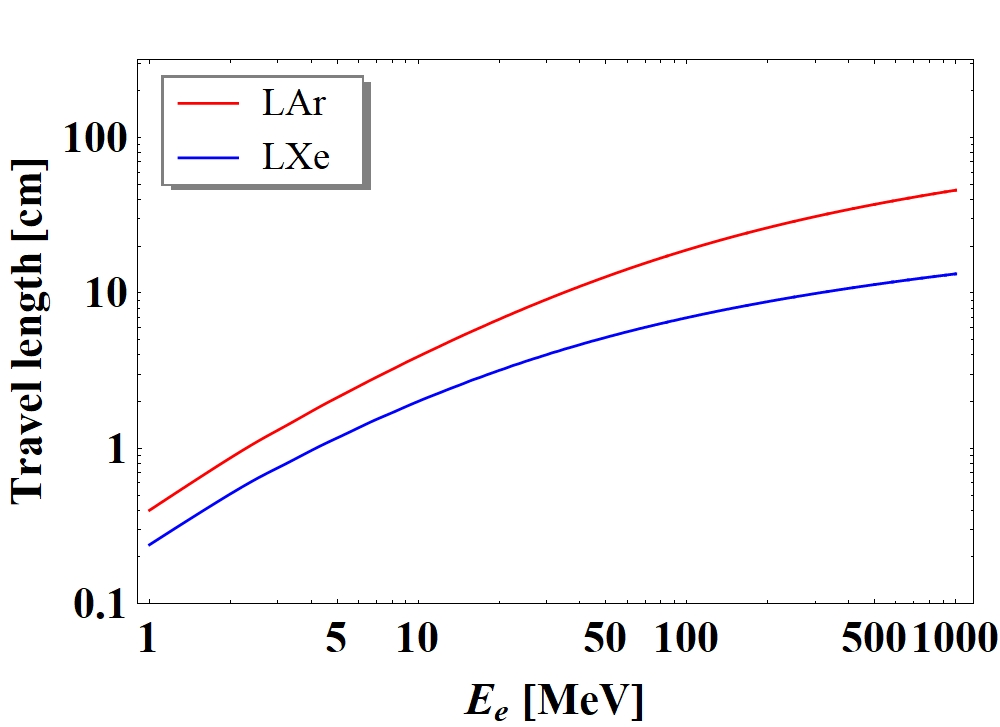}
\caption{\label{fig:stop} Stopping power (left panel) and travel length (right panel) for electrons in LXe and LAr detectors as a function of the electron energy, based on data available in~\cite{NIST}, for densities of LXe and LAr of 3 g/cm$^3$ and 1.5 g/cm$^3$. }
\end{figure}

First of all, we claim that electrons in our signals will  travel an appreciable distance in the detector. The left panel of Fig.~\ref{fig:stop} exhibits the stopping power for electrons in the LXe (red) and LAr (blue) detectors as a function of the electron energy. These results are obtained from data available in~\cite{NIST}, using LXe and LAr densities of 3 g/cm$^3$ and 1.5 g/cm$^3$, respectively. To calculate the travel length as a function of the electron energy, we integrate the change in stopping power; our results are shown in the right panel of Fig.~\ref{fig:stop}. We expect that an electron with energy $\sim$ 10--1000 MeV will travel $\sim$ 2--10 ($\sim$ 4--40) cm before it stops in the LXe (LAr) detector.

Since energetic electron signals are expected to give sizable tracks, their energy deposits are highly non-trivial. We present our predictions in terms of two extreme cases: purely vertical (i.e., in the $z$ direction) and purely horizontal (i.e., in the $xy$ plane) tracks.
For the former case, free electrons will be drifted upward and deposit energy as a form of S2 from the electrons closer to the gaseous Xe to those further, in order.
Since the time resolution of PMTs is as good as $\sim 10$ ns~\cite{Aprile:2017aty}, we expect that S2 will appear as a series of flickers of several PMTs, and thus the energy and track reconstruction will be sufficiently good~\cite{com}.
When it comes to the latter case, we expect that many free electrons will create S2 scintillation photons simultaneously within a single cycle of the PMT measurement so that some of the PMTs may saturate. 
Nevertheless, we also expect that there will be unsaturated PMTs with which the energy and the track are reconstructed with reasonable precision~\cite{com}.
In addition, an energetic signal may allow S1 itself to develop informative patterns among PMTs, which can be also used for energy and/or track reconstruction.

We point out that a positron can manifest itself as a distinguishable Bragg peak on top of all features produced by electron signals.
Once a positron stops in the detector, it will annihilate with a nearby electron, creating two characteristic photons with energy of 511 keV.
Therefore, the detection of such a line at the tail of a track could be considered as strong evidence that the observed signal differs not only from background but also from elastic DM signal.
This signature is particularly useful for DEAP-3600 as PMTs cover almost the entire surface of the detector.

In summary, a signal via the DM-electron scattering will be featured by energetic, appreciable track(s). 
For the case of inelastic scattering, the (reconstructed) interaction points may be displaced from each other. 
If the incident $\chi_1$ is too boosted, the associated tracks may overlap, rendering track identification somewhat challenging.
However, we expect that multiple tracks will be resolved/identified fairly well, especially due to great resolution in the $z$ direction and the 511 keV peak.

Finally, we emphasize that an energetic, sizable track is sufficient for claiming a discovery, because of its uniqueness. 
Additional features including multiple tracks are more relevant to post-discovery studies such as the disambiguation of underlying dynamics.

\subsection{Background consideration}

In this section, we discuss potential background sources to our signal events. As standard non-relativistic DM searches at conventional DM direct detection experiments are based on the ``zero'' background assumption, we investigate whether this assumption is still valid in the relativistic DM-electron scattering, and if so, under what conditions.

As argued in the previous section, a typical signal event will involve energetic track(s) that can occur within the fiducial volume. 
Potential candidate backgrounds producing similar features could be atmospheric neutrinos and solar neutrinos, especially from $^8$B.
The flux of the former is too small to be observed even with a 5-year run of LZ. However, we find that the contribution from the latter class of background cannot be simply neglected, as the total flux of the $^8$B solar neutrino is $\sim 10^6$ cm$^{-2}$s$^{-1}$ in the $\sim$ 1--15 MeV range~\cite{Billard:2013qya}.
The scattering cross section for $^8$B solar neutrinos incident on electrons is thoroughly investigated for different values of the minimum accepted kinetic energy in Ref.~\cite{Bahcall:1986pf}.
We estimate that, with an acceptance cut on ER events with energy greater than 10 MeV,  the expected number of background events from $^8$B solar neutrino is smaller than one, even at LZ-5year.

To ensure the zero background assumption, we define our signal region with an energy cut of 10 MeV and with at least a single track within the fiducial volume.\footnote{Of course, a 10 MeV cut is excessively severe for XENON1T and DEAP-3600, given that their detector fiducial volume is smaller than for LZ. For simplicity, we take a common energy cut so our results can be regarded as conservative.}
A possible issue comes from the energy resolution in the relevant energy regime because of the potential presence of saturated PMTs. 
We nevertheless expect an $\mathcal{O}(10\%)$ accuracy in the energy determination using information from unsaturated PMTs~\cite{com}, and the background is still negligible even if the energy cut is reduced to 9 MeV.

\subsection{Phenomenology}

\begin{table*}
\begin{tabular}{c  c  c  c  c  c  c  c  c  c }
 & & \multicolumn{2}{c}{ref1} & \multicolumn{2}{c}{ref2} & \multicolumn{2}{c}{ref3} & \multicolumn{2}{c}{ref4}  \\
\hline
\multicolumn{2}{c}{Expected flux} & \multicolumn{2}{c}{610} & \multicolumn{2}{c}{43} & \multicolumn{2}{c}{0.98} & \multicolumn{2}{c}{0.24} \\
\hline
Experiments & ~~Run time~ & ~~~~~multi~ & ~single~ & ~~~multi~ & ~single~ & ~~~multi~ & ~single~ & ~~~multi~ & ~single  \\
\hline
\multirow{2}{*}{XENON1T} &  1yr  & ~~~~~2000~ & ~160~ & ~~~220~ & ~7.5~ & ~~~0.37~ & ~0.37~ & ~~~0.27~ & ~0.27 \\
 & 5 yr & ~~~~~~390 & ~32 & ~~43~ & ~1.5~ & ~~~0.075~ & ~~0.075~ & ~~~0.054~ & ~0.054 \\
\hline
\multirow{2}{*}{DEAP-3600} & 1 yr & ~~~~~~450 & ~63~  & ~~~55~ & ~3.1~ & ~~~--~ & ~0.16~ & ~~~--~ & ~0.11 \\
 & 5 yr & ~~~~~~~91 & ~13~ & ~~~11~ & ~0.61~ & ~~~--~ & ~0.031~ & ~~~--~ & ~0.022 \\
\hline
\multirow{2}{*}{LZ} & 1 yr & ~~~~~~180 & ~27~ & ~~~25~ & ~1.3~ & ~~~0.067~ & ~0.067~ & ~~~0.048~ & ~0.048 \\
 & 5 yr & ~~~~~~~36 & ~5.4~ & ~~~5.0~ & ~0.26~ & ~~~0.013~ & ~0.013~ & ~~~0.0096~ & ~0.0096 \\
\end{tabular}
\caption{\label{table:fluxexp} For each reference point, defined in the table in Fig.~\ref{fig:spectrum}, we give (on the first line) the expected flux of $\chi_1$ over the whole sky calculated under the assumption of $\langle \sigma v \rangle_{0\rightarrow 1} = 5\times 10^{-26}$ cm$^3$s$^{-1}$ at the present time, and (on the rest of the table) the flux of $\chi_1$ required for detection via inelastic scattering. All fluxes are given in unit of 10$^{-3}$ cm$^{-2}$s$^{-1}$. 
Two cases are considered: ``multi'', corresponding to multiple tracks in the fiducial volume and ``single'', corresponding to at least one single track with energy larger than 10 MeV.
}
\end{table*}

Equipped with the detection strategies and background considerations discussed in the preceding sections, we can now study the phenomenology of iBDM and the experimental discovery reach. We start by estimating the flux needed to reach experimental sensitivity, for a given period of exposure and for the reference points defined in the table in Fig.~\ref{fig:spectrum}.
As before, the relevant Monte Carlo simulation is performed using \texttt{MG5@aMC}~\cite{Alwall:2014hca}.
For a given event in each reference point, we first determine the primary scattering point and the incident angle by a set of random number generation, and then shift/rotate all particle momenta in the event accordingly.
We then calculate the decay distance of $\chi_2$ or $X$ according to the exponential decay law.
If the interaction points for the three $e^{\pm}$'s are located inside the detector fiducial volume, we tag the event as accepted.
However, if the secondary vertex lies outside the fiducial volume, we accept the events with primary recoil electron energy greater than 10 MeV.
For each reference point, we generate 10,000 events and compute the associated acceptance in combination with the number of accepted events.

We emphasize that one to three energetic (isolated) $e^-$ and/or $e^+$ signals passing the above-mentioned selection criteria are considered unique enough not to be faked by potential backgrounds. Hence, we quantify the experimental sensitivity by demanding 2.3 DM signal events which correspond to the 90\% C.L. upper limit under the assumption of a null observation over a zero background with Poisson statistics.
Table~\ref{table:fluxexp} reports the minimum required fluxes of $\chi_1$ allowing our reference points to be sensitive at the experiments we consider within given periods of exposure.
The decay widths are obtained numerically, without relying on the approximation in Eq.~\eqref{eq:app}.
Two channels are considered: {\it i)} ``multi'', corresponding to multiple tracks in the fiducial volume and {\it ii)} ``single'', corresponding to at least one single track with energy larger than 10 MeV.
The latter is more inclusive than the former, unless an event does not pass the energy cut but gives detectable multiple tracks.  In our selection
we did not include events in which the primary process takes place outside the fiducial volume, but the secondary process occurs within and can be detected. Including these events would improve the signal sensitivity.
Note that DEAP-3600 has no sensitivity to ref3 and ref4 for the multi-signal channel. This is because the dark photon $X$ in both points decays rather promptly, whereas for DEAP-3600 the position resolution is greater than 6.5 cm and the S2 mode is unavailable.
For comparison, we also present the expected flux of $\chi_1$ over the whole sky for each reference point assuming that $\langle \sigma v \rangle_{0\rightarrow1}$ (the velocity-averaged cross section of $\chi_0\chi_0\rightarrow \chi_1\chi_1$) is $5\times 10^{-26}$ cm$^3$/s at the present time.\footnote{We here assume the NFW halo profile as in Ref.~\cite{Agashe:2014yua}.}
Our conclusion is that all reference points are quite promising for signal detection. 

Next we study the model-{\it independent} experimental reach in parameter space, in the spirit of counting experiments. The number of signal events $N_{\text{sig}}$ is given by 
$
N_{\text{sig}} = \sigma  \, \mathcal{F} \, A \, t_{\textrm{exp}} \, N_e \, ,
$
where $\sigma$ is the cross section for $\chi_1 e \rightarrow \chi_2 e$, $\mathcal{F}$ is the $\chi_1$ flux, $A$ is the acceptance, $t_{\textrm{exp}}$ is the exposure time, and $N_e$ is the number of electrons in target material. We are assuming for simplicity that the branching ratio of $\chi_2 \rightarrow \chi_1 e^+e^-$ is equal to one. 
The last two parameters ($t_{\textrm{exp}}$ and $N_e$) are determined solely by the characteristics of the experiment. On the contrary, the product $\sigma \, \mathcal{F}$ is a function of all model parameters such as masses and coupling constants. Finally, the acceptance $A$ (defined as 1, if the event occurs within the fiducial volume, and 0 otherwise) is determined by the distance between primary and secondary vertices $\ell_{\textrm{lab}}$.

We determine the experimental sensitivity by demanding at least 2.3 signal events, which correspond to the inequality
\bea
\sigma \, \mathcal{F} > \frac{2.3}{A(\ell_{\textrm{lab}})\, t_{\textrm{exp}} \, N_e}\,. \label{eq:sense}
\eea
The left-hand side of Eq.~(\ref{eq:sense}) is a model-dependent quantity, while the right-hand side depends only on experimental specifications and on $\ell_{\textrm{lab}}$, which is different event-by-event. Thus, we think that the best way to express the experimental sensitivity (and, in the future, experimental results) is in the plane $\ell_{\textrm{lab}}^{\max}$--$\sigma \mathcal{F}$, where $\ell_{\textrm{lab}}^{\max}$ is the maximum mean decay length of a long-lived particle, i.e., either $\chi_2$ or $X$.

\begin{figure}[tbp]
\includegraphics[width=8.2cm]{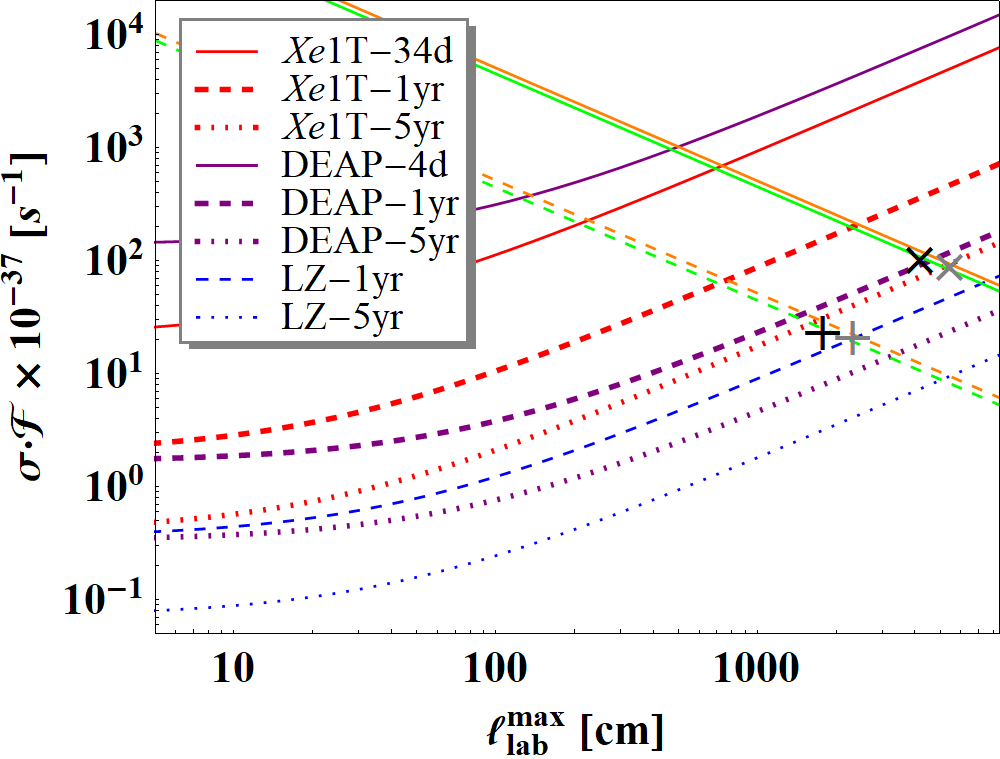} \\
\includegraphics[width=8.2cm]{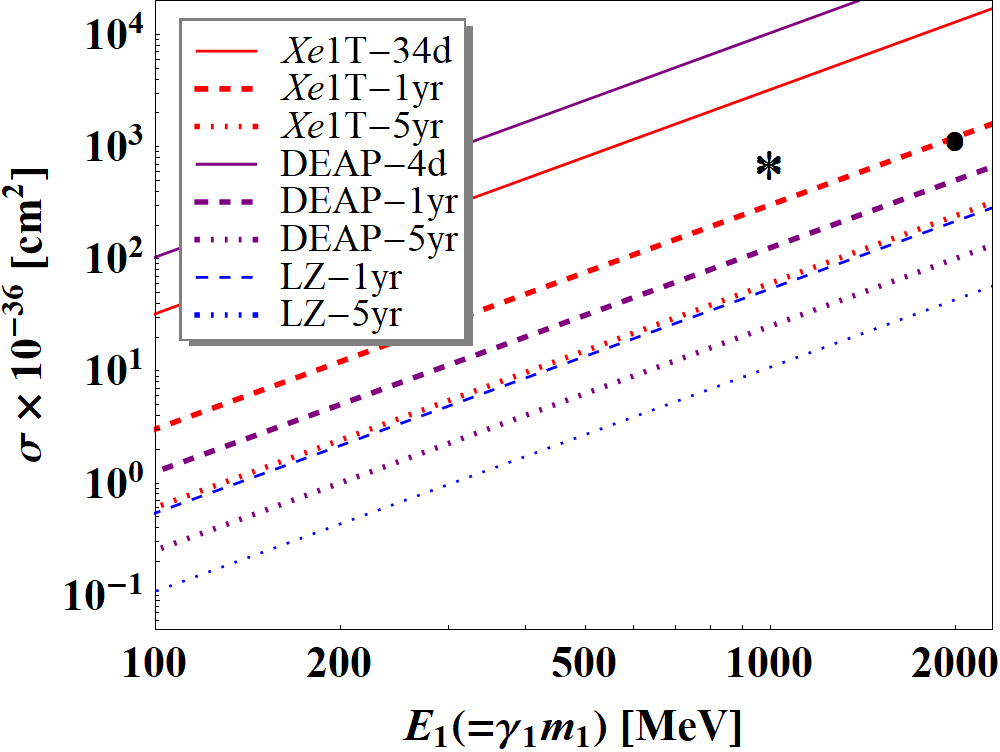}
\caption{\label{fig:sigmaSI} Top: Experimental sensitivities in the $\ell_{\textrm{lab}}^{\max}$--$\sigma \mathcal{F}$ plane at XENON1T-34.2days~\cite{Aprile:2017iyp}, XENON1T-1yr, XENON1T-5yr, DEAP-3600-4.44days~\cite{Amaudruz:2017ekt}, DEAP-3600-1yr, DEAP-3600-5yr, LZ-1yr, and LZ-5yr under zero-background assumption, for the case of displaced secondary vertices. Bottom: Experimental sensitivities in the $E_1$--$\sigma$ plane for the case of prompt decays.
We take $\langle \sigma v \rangle_{0\rightarrow 1} = 5\times 10^{-26}$ cm$^3$s$^{-1}$ to evaluate the $\chi_1$ flux from Eq.~(\ref{eq:flux}).
The four symbols $\times$, $+$, $*$, $\bullet$ locate our reference points ref1, ref2, ref3, ref4, respectively (with grey symbols for the case of scalar DM). The green/orange solid/dashed lines show the predictions of fermion/scalar DM with ref1/ref2 mass spectra and varying $\epsilon$. 
}
\end{figure}

The top panel of Fig.~\ref{fig:sigmaSI} shows the experimental reach of our benchmark detectors under the assumption of cumulatively isotropic $\chi_1$ flux. In the same figure, we also show the predictions for our reference points ref1 (indicated in black by the symbol $\times$) and ref2 (symbol $+$), with the flux $\mathcal{F}$ computed as in Eq.~(\ref{eq:flux}) by taking $\langle \sigma v \rangle_{0\rightarrow 1} = 5\times 10^{-26}$ cm$^3$s$^{-1}$. Our results suggest that both reference points are within the discovery reach with 5-year runs of XENON1T or DEAP-3600. This conclusion is consistent with the numbers reported in Table~\ref{table:fluxexp}. The grey symbols $\times$ and $+$ refer to the parameter choice of ref1 and ref2, but with the DM particle being scalar, instead of fermion. Because of the smaller cross section and decay width for scalar DM, the sensitivity to scalar DM is slightly worse than for fermion DM. As $\sigma \mathcal{F}$ and $\ell_{\text{lab}}^{\max}$ are respectively proportional to $\epsilon^2$ and $1/\epsilon^2$, we can easily compute the effect of varying the coupling $\epsilon$, while keeping fixed all mass parameters of our reference points. This is shown by the green (orange) solid and dashed, which correspond to the mass spectra of ref1 and ref2, for fermion (scalar) DM model, as the parameter $\epsilon$ is varied; $\epsilon$ becomes larger (smaller) towards the upper-left (lower-right).

If the secondary process is prompt, the analysis is simplified as the acceptance $A$  becomes equal to one. 
Since $\mathcal{F}$ is inversely proportional to $m_0^2$ (or, equivalently, to $E_1^2$) for fixed $\langle \sigma v \rangle_{0\rightarrow 1} = 5\times 10^{-26}$ cm$^3$s$^{-1}$, the most appropriate way to visualize the experimental sensitivity given in Eq.~(\ref{eq:sense}) is by
presenting our results in a $E_1$--$\sigma$ plane.

The bottom panel of Fig.~\ref{fig:sigmaSI} shows the corresponding experimental reach in our benchmark experiments, in the case of prompt decays, under the same assumptions as before. We point out that this experimental sensitivity is relevant to models with elastic scattering as well, in combination with a proper energy cut to satisfy the zero-background assumption, since it does not specify the number of final-state particles.
While we have assumed $\langle \sigma v \rangle_{0\rightarrow 1} = 5\times 10^{-26}$ cm$^3$s$^{-1}$, our results can be easily rescaled to different values of the $\chi_1$ flux in the present universe, originating from $\chi_0$ self-annihilation, $\chi_0$ decay, or other sources. The predictions of the reference points ref3 and ref4 are indicated by the symbols $*$ and $\bullet$, with the results for the fermion and scalar models almost overlapping.

We make a couple of comments on the experimental sensitivity at DEAP-3600. 
Since its position resolution does not allow for identification of multiple particles without any appreciable displaced vertex, we trigger a signal observation by requiring only a single ER inside the fiducial volume.
This raises an issue, which is also relevant to the case of elastic scattering. As discussed before, the zero-background assumption for the single-track signal requires a suitable energy cut to eliminate the contamination from $^8$B solar neutrinos. Therefore, the sensitivities for low $E_1$ would be slightly degraded as the cut affects the relevant signal acceptance.

\begin{figure}[t]
\includegraphics[width=0.4944\linewidth]{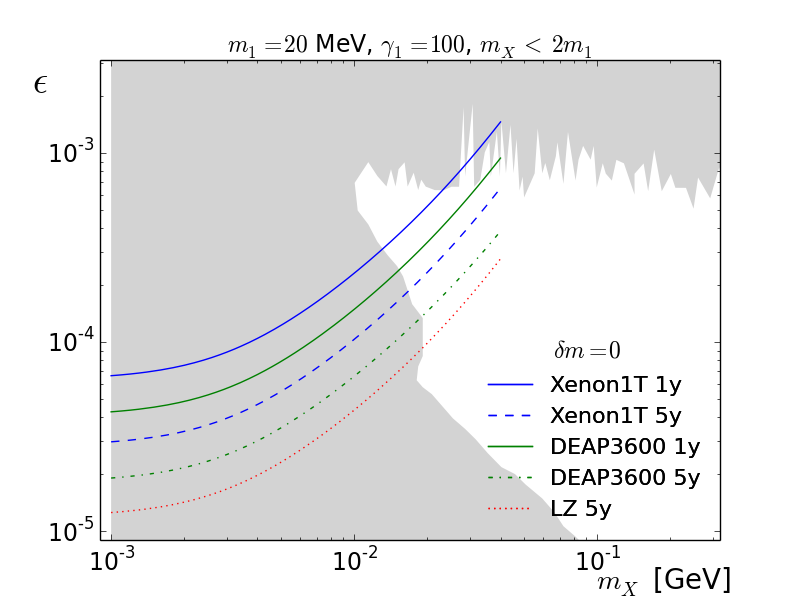}
\includegraphics[width=0.4944\linewidth]{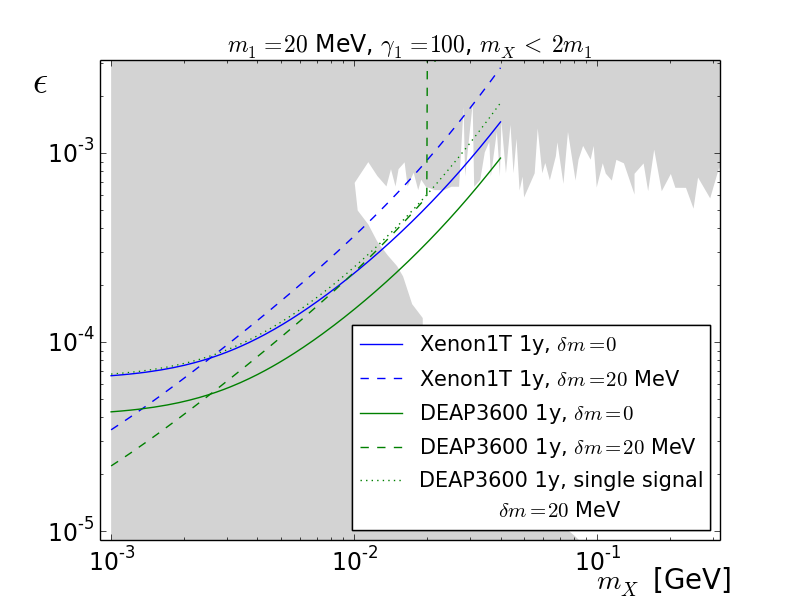}
\includegraphics[width=0.4944\linewidth]{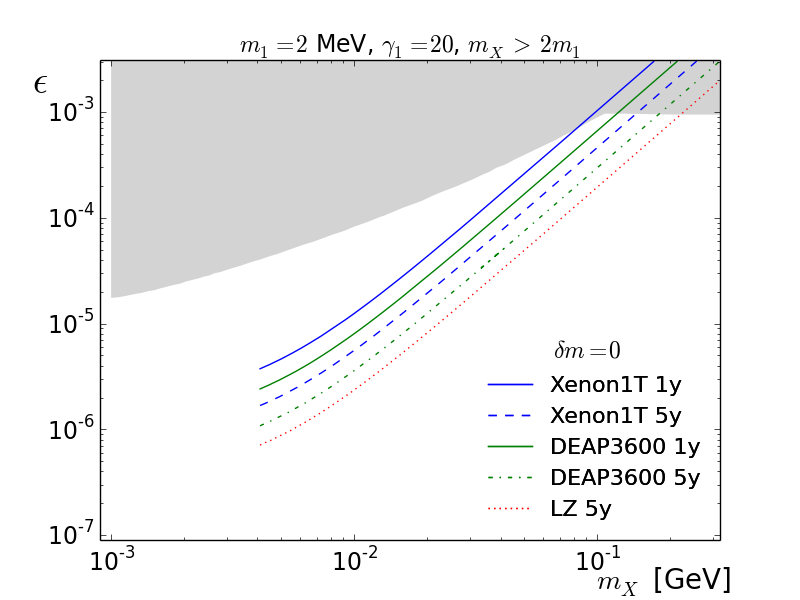}
\includegraphics[width=0.4944\linewidth]{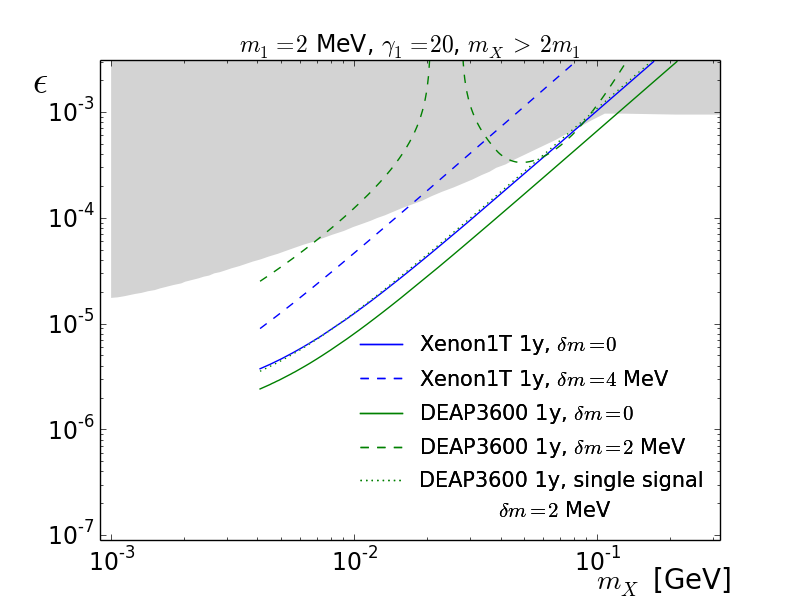}
\caption{\label{fig:mXepsilon}
Experimental reach at various experiments in the $m_X$--$\epsilon$ plane for the case in which the dark photon $X$ decays visibly (top panels) or invisibly (bottom panels). The grey regions show the currently excluded parameter space, as reported in Refs.~\cite{Essig:2013lka} (top panels) and \cite{Banerjee:2017hhz} (bottom panels).
The left panels show the results of elastic scattering at different experiments, while in the right panels we compare cases of elastic ($\delta m =0$) and inelastic ($\delta m \ne 0$) scattering. 
}
\end{figure}

It is interesting to relate our results with other analyses of dark photon phenomenology, which are usually expressed in the $m_X$--$ \epsilon$ plane. 
To this end, we fix $E_1$ (or, equivalently, $m_0$), $m_1$, $m_2$, and the dark-sector gauge couplings (taking $g_{11}=1$ for elastic and $g_{12}=1$ for inelastic scattering), and  find the minimum value of $\epsilon$ for a given $m_X$ while scanning all boundary points in $(\ell_{\textrm{lab}}^{\max},\sigma \mathcal{F})$.
Our results are displayed in Fig.~\ref{fig:mXepsilon}.
We assume that the dark photon $X$ decays either visibly (top panels) or invisibly (bottom panels), and choose the mass spectra accordingly. The invisible decay can be  $X\to \chi_1 \chi_1$ (which dominates $X\to \chi_1 \chi_2$ when the phase-space factor wins over the ratio of gauge couplings $g_{12}^2/g_{11}^2$) or into other long-lived states present in the dark sector.
We also show the current excluded parameter regions (grey areas) obtained from Refs.~\cite{Essig:2013lka} (top panels) and \cite{Banerjee:2017hhz} (bottom panels).

In the left panels of Fig.~\ref{fig:mXepsilon}, we report the sensitivities in various experiments for the elastic scattering scenario.
We fix $(E_1, m_1)$ to be (2 GeV, 20 MeV) and (40 MeV, 2 MeV) for top-left and bottom-left panels, respectively.
We see that searches in the elastic scattering channel allow for explorations of parameter regions that are not covered by existing experimental constraints. 
In particular, for the case of invisibly decaying dark photon (bottom-left), XENON1T and DEAP-3600 can probe values of $\epsilon$  one order of magnitude below current bounds.

In the right panels of Fig.~\ref{fig:mXepsilon}, we show some cases with parameters corresponding to inelastic scattering scenarios ($\delta m \ne 0$) and compare them with cases of elastic scattering ($\delta m = 0$). 
Since the cross section for the elastic scattering is larger than for the inelastic channel (at equal gauge couplings),
typically the experimental reach of the latter is slightly worse. However, when we consider smaller mass regions, the elastic signal suffers from the energy cut $E_e>10$~MeV, while the inelastic signal remains unaffected as it can be detected through multi-track events. As a result, the sensitivity on the inelastic scattering can be better than the ordinary case of elastic scattering.
However, at DEAP-3600, it is more challenging to tag signal events by requiring multiple final-state particles, if their interaction points are displaced by a distance smaller than the position resolution of DEAP-3600 (see the green dashed curve in the bottom-right panel). 
More interestingly, there is a competing effect between the scattering cross section and the decay length. Indeed, as $m_X$ increases, the cross section becomes smaller while the decay length of $\chi_2$ becomes larger. One can relax the selection criteria by requiring a single-track signal together with a proper energy cut mentioned earlier in order to achieve much better sensitivities even at DEAP-3600 (see, for example, the green dotted curve in the bottom-right panel, which closely follows the blue solid curve).

In this paper, we have concentrated our study on how direct detection experiments can explore both iBDM  and the elastic channel of boosted DM.  However, we point out that the same class of DM models can also be probed by different kind of experiments, such as:
\begin{itemize} \itemsep0em
\item Solar neutrino detectors, e.g., Borexino~\cite{Alimonti:2008gc, Back:2012awa} which comes with a great energy resolution and a larger fiducial volume.
\item Array-type detectors in neutrinoless double beta decay experiments, e.g., CUORE~\cite{Alduino:2016vjd, Alduino:2017ehq}, and in DM direct detection experiments, e.g., COSINE-100~\cite{Adhikari:2017esn}, which are particularly good at identifying displaced vertices. 
\end{itemize}


\section{Conclusions} \label{sec:outlook}

In this paper we have explored a new class of DM models. In the context of multi-particle dark sectors, it is possible to combine the properties of {\it inelastic} DM~\cite{TuckerSmith:2001hy} and {\it boosted} DM~\cite{Agashe:2014yua} and construct models of iBDM~\cite{Kim:2016zjx}. The interest in these models lies in their radically different signals that can be tested in ordinary DM direct detection experiments, using unconventional search criteria. 

We have argued that highly boosted MeV-range DM can be probed at conventional direct detection experiments via relativistic scattering of electrons inside target material. Taking a broad and fairly model-independent approach, we have considered both the cases of elastic scattering off electrons (as for boosted DM) and inelastic scattering (which is the distinguishing feature of iBDM).
For iBDM, the signal is characterized by energetic recoil electrons accompanied by additional visible particles with distinctive vertex displacements. The peculiarity of the signal allows for searches in a background-free environment. For elastic scattering, an acceptance cut in the electron recoil energy of about 10 MeV is sufficient to reject backgrounds from atmospheric and solar neutrinos. These results motivate the search for energetic electrons in DM direct detection experiments as a novel way to probe unconventional dark sectors. 

We have examined
detection prospects in the context of dark photon scenarios at XENON1T, DEAP-3600, and LUX-ZEPLIN detectors, considering both fermion and scalar DM and discussing the expected signal features. We have found that ordinary DM direct detection experiments have sufficient sensitivity to the electron recoil signatures to probe unexplored regions of the underlying models, and DM events might be observed even collecting only one-year of data at XENON1T. In Fig.~\ref{fig:sigmaSI} we have shown a model-independent way in which future experimental results can be plotted and conveniently confronted with specific theoretical predictions of iBDM scenarios.  
We have also re-expressed the experimental reach of our proposed DM searches in terms of the parameter space often used for dark photon analyses (see Fig.~\ref{fig:mXepsilon}). Our results show that DM detection experiments can explore regions of dark photon models that are left uncovered by other experimental technics. We have also pointed out that neutrino detectors and array-type detectors searching for neutrinoless double beta decay are potentially sensitive to the DM signals that we have suggested.

Given our promising findings and results, we strongly encourage relevant experimental collaborations to revisit past analyses and perform future studies in the search for the novel DM signatures proposed in this paper.

\section*{Acknowledgments}
We thank Yang Bai, Joshua Berger, Walter Bonivento, Bogdan Dobrescu, Patrick Fox, Andr\'{e} de Gouv\^{e}a, Roni Harnik, Ahmed Ismail, Gordan Krnjaic, Maxim Pospelov, Richard Ruiz, and Yue Zhang for insightful discussions and Olivier Mattelaer for help with simulation.
We particularly acknowledge Eric Dahl, Luca Grandi, Yeongduk Kim, Rafael Lang, Hyunsu Lee, and Aaron Manalaysay for dedicated discussions on the experimental aspects.
This work was performed in part at the Aspen Center for Physics, which is supported by National Science Foundation grant PHY-1607611.
DK and JCP acknowledge the Galileo Galilei Institute for Theoretical Physics for the hospitality and the INFN for partial support during the completion of this work.
DK, JCP and SS also appreciate the hospitality of Fermi National Accelerator Laboratory.
DK is supported by the Korean Research Foundation (KRF) through the CERN-Korea Fellowship program.
JCP is supported by the National Research Foundation of Korea (NRF-2016R1C1B2015225) and the POSCO Science Fellowship of POSCO TJ Park Foundation.
SS is supported by the National Research Foundation of Korea (NRF-2017R1D1A1B03032076).

\appendix
\section{}
\numberwithin{equation}{section}

Here we will present some explicit examples of dark photon models of iBDM, emphasizing how off-diagonal interactions can dominate over diagonal interactions.

\bigskip

\paragraph*{{\bf Model 1: Two chiral fermions.}}
Consider two fermions ($\psi_L$ and $\psi_R$) of opposite chirality and same charge under a dark U(1)$_X$.
The most general Lagrangian includes Majorana and Dirac mass terms
\bea
- {\mathcal L}_m &=& m_L \psi_L^T C \psi_L + m_R \psi_R^T C \psi_R + 2m_D {\bar \psi}_R\psi_L + {\rm h.c.} \nonumber\\
&=& \Psi^TCM\Psi + {\rm h.c.}
\eea
\beq
\Psi \equiv \begin{pmatrix}\psi_L\\ \psi_R^c \end{pmatrix}\, , \qquad
M\equiv \begin{pmatrix}m_L & m_D \cr m_D & -m_R\end{pmatrix} \, ,
\eeq
where, for simplicity, the parameters $m_{L,R,D}$ are taken to be real. The Majorana mass term is generated by effective interactions after spontaneous U(1)$_X$ breaking.
The mass eigenstates are two left-handed chiral fermions $\chi_{1,2}$ such that
\beq
- {\mathcal L}_m = \sum_{i=1}^2 m_{\chi_i} \chi_i^T C \chi_i+ {\rm h.c.}
\eeq
\beq
\begin{pmatrix} \chi_2 \\ \chi_1 \end{pmatrix} = \begin{pmatrix} \cos \theta & \sin\theta \\ -\sin\theta & \cos\theta \end{pmatrix}
\begin{pmatrix}\psi_L \\ \psi_R^c \end{pmatrix} \, , \quad
\tan 2\theta = \frac{2m_D}{m_L+m_R} \, ,
\eeq
\beq
m_{\chi_{2,1}}=\frac{1}{2}\left( m_L-m_R \pm \sqrt{(m_L+m_R)^2+4m_D^2}\right) \, .
\eeq

The gauge interactions are defined by the Lagrangian
\bea
- {\mathcal L}_g &=& \frac14 F_{\mu \nu}F^{\mu\nu} +\frac14 X_{\mu \nu}X^{\mu\nu} +\frac{\epsilon}{2} X_{\mu \nu}F^{\mu\nu}
-\frac{m_X^2}{2}X_\mu X^\mu \nonumber \\
&&+ eA_\mu J_{\rm em}^\mu +g_X X_\mu J_X^\mu \, ,
\label{lag1}
\eea
where $A_\mu$ is the ordinary photon with field strength $F_{\mu\nu}$ and electromagnetic current $J_{\rm em}^\mu$, while $X$ is the dark photon with field strength $X_{\mu\nu}$ and gauge coupling $g_X$.
The small parameter $\epsilon$ measures the kinetic mixing between dark and ordinary photon.
It is convenient to rewrite \eq{lag1} by making the shift $A_\mu \to A_\mu -\epsilon X_\mu$ and working at the leading order in $\epsilon$
\bea
- {\mathcal L}_g &=& \frac14 F_{\mu \nu}F^{\mu\nu} +\frac14 X_{\mu \nu}X^{\mu\nu}
-\frac{m_X^2}{2}X_\mu X^\mu \nonumber \\
&&+ eA_\mu J_{\rm em}^\mu +X_\mu \left( g_X J_X^\mu -e\epsilon J_{\rm em}^\mu\right)  \, .
\label{lag2}
\eea
The kinetic terms are now canonical, but the dark photon current has gained an additional term proportional to the electromagnetic current.

The new chiral fermions contribute to the dark photon current as
\beq
J_X^\mu ={\bar \psi}_L \gamma^\mu \psi_L + {\bar \psi}_R \gamma^\mu \psi_R= {\bar \psi}_L \gamma^\mu \psi_L - {\bar \psi}^c_R \gamma^\mu \psi_R^c \, ,
\eeq
which, in terms of mass eigenstates, becomes
\beq
J_X^\mu =\cos 2\theta \left( {\bar \chi_2} \gamma^\mu \chi_2 -   {\bar \chi_1} \gamma^\mu \chi_1 \right) -\sin 2\theta \left( {\bar \chi_2} \gamma^\mu \chi_1  + {\rm h.c.}\right) \, .
\eeq
These expressions can be matched to the simplified model of Eq.~\eqref{eq:lagrangian}, with the result $g_{12}/g_{11} =\tan 2\theta$.
The $J_X^\mu$ current contains both diagonal and off-diagonal components and thus the dark photon exchange contributes to both elastic and inelastic scattering processes. Off-diagonal couplings dominate over diagonal couplings for $2m_D \gg m_L +m_R$. This regime does not necessarily imply small mass differences between $\chi_2$ and $\chi_1$.
For instance, once we take $m_R \approx -m_L$ we find $\sin 2\theta \approx 1$ and $m_{\chi_{2,1}} \approx |m_D \pm m_L |$, so the mass splitting can be arbitrarily large even for vanishingly small diagonal couplings.
Note that, for sufficiently heavy dark photon, all model dependence can be lumped into a single parameter ($\epsilon g_X\sin 2 \theta /m_X^2$) that describes the effective interaction governing both the primary and secondary processes.

\bigskip

\paragraph*{{\bf Model 2: Two real scalars.}}
Consider a complex scalar $\phi$ charged under the dark U(1)$_X$.
The most general mass term is
\beq
- {\mathcal L}_m =m_\phi^2 |\phi |^2 + \frac{\delta}{2} (\phi^2 + {\rm h.c.}) \, .
\label{lag3}
\eeq
The mass term proportional to $\delta$ comes from the spontaneous breaking of $U(1)_X$, which is also responsible for the mass $m_X$ of the dark photon.
We rewrite \eq{lag3} in terms of two real scalars $\chi_{1,2}$, which are mass eigenstates,
\beq
- {\mathcal L}_m =\frac12 \sum_{i=1}^2 m_{\chi_i}^2 \chi_i^2 \, , ~~
\phi \equiv \frac{\chi_2 +i \chi_1}{\sqrt{2}} \, , ~~
m^2_{\chi_{2,1}}=m_\phi^2\pm \delta \, .
\eeq

The interaction Lagrangian is the same as in \eq{lag2}, where the contributions of the two scalars to the dark photon current is
\beq
J_X^\mu = i \left( \phi^* \partial^\mu \phi - \phi \partial^\mu \phi^* \right) = \chi_1 \partial^\mu \chi_2 -\chi_2 \partial^\mu \chi_1 \, .
\eeq
Unlike the fermion case, we find that the scalar current is purely off-diagonal. This is because the scalars $\chi_2$ and $\chi_1$ have opposite CP parity and, as long as CP is conserved, one cannot construct a current involving only a single state.
As a result, the dark photon can mediate inelastic primary processes, but no elastic scattering.
In this case, the single parameter that describes the effective interaction, valid for sufficiently heavy dark photon, is $\epsilon g_X  /m_X^2$.

\section{}

\paragraph*{{\bf Scattering cross section.}} 

The matrix element squared for the process $\chi_1 e \rightarrow \chi_2 e$  is given by~\cite{Kim:2016zjx}
\bea
\overline{\left|\mathcal{M} \right|}^2
= \frac{(\epsilon e g_{12})^2}{[2m_e(E_{2}-E_{1})-m_{X}^2]^2}\mathcal{M}_0\,,  \label{eq:matrixX}
\eea
where $E_i$ ($i=1,2$) is the $\chi_i$ energy measured in the laboratory frame. We find that $\mathcal{M}_0$ for our benchmark models are, for scalar and fermion, respectively
\bea
\mathcal{M}_{0,s} &=& 8m_e\left(2m_e E_1E_2 +m_1^2E_2-m_2^2 E_1 \right)\,, \\
\mathcal{M}_{0,f} &=& 8m_e\left[m_e( E_{1}^2+E_{2}^2)-\frac{(\delta m)^2}{2}(E_{2} -E_{1} +m_e)\right. \nonumber \\
&&+\left. m_e^2(E_{2}-E_{1})+m_{1}^2E_{2} - m_{2}^2 E_{1} \right]\,.
\eea
The differential cross section in the laboratory frame is simply given by
\begin{align}
\frac{d\sigma}{dE_e}=\frac{m_e }{8\pi\lambda^2(s,m_e^2,m_{1}^2)}\overline{\left|\mathcal{M} \right|}^2 \,, \label{eq:ETspec}
\end{align}
where $E_e$ is the total energy of the recoiling electron and $\lambda(x,y,z) = [(x - y - z)^2 - 4yz]^{1/2}$.
Here $\overline{\left|\mathcal{M} \right|}^2$ is written in terms of $E_{e} = E_{1}+m_e-E_{2}$ and $s=m_1^2+m_e^2+2m_eE_1$.

\bigskip

\paragraph*{{\bf 3-body decay width.}} 
The differential $\chi_2$ decay width for $\chi_2(p_2)\rightarrow \chi_1(p_1) e^-(p_{e^-}) e^+(p_{e^+})$ is
\bea
&&\!\!\!\!\!\!\!\!\!\!\!\!\!\!\!\!\!\!
 \frac{d^2\Gamma_2}{dE_{e^-} dE_{e^+}} =\frac{g_{12}^2\epsilon^2 \alpha}{16\pi^2 m_2 }\times \nonumber \\
&&\!\!\!\!\!\!\!\!\!
 \frac{\Theta(1-\cos^2\theta_{ee})\overline{|\mathcal{A}|}^2}{[m_2^2-2m_2 E_1+m_1^2-m_X^2]^2+m_X^2\Gamma_X^2} 
 \,,
\eea
where $\Theta(x)$ denotes the usual Heaviside step function and $\cos\theta_{ee}$ is given by
\bea
\cos\theta_{ee} = \frac{E_1^2-P_{e^-}^2-P_{e^+}^2-m_1^2}{2P_{e^-}P_{e^+}}\,,
\eea
with $P_{e^{\pm}}^2=E_{e^{\pm}}^2-m_e^2$. We work in the $\chi_2$ rest frame, where the following relations hold 
\bea
E_{e^-}=\frac{m_2^2+m_e^2-s_2}{2m_2}\, ,~~ E_{e^+}=\frac{m_2^2+m_e^2-s_1}{2m_2}
 \label{eq:epem}
\eea
with $s_1=(p_1+p_{e^-})^2$ and $s_2=(p_1+p_{e^+})^2$. The spin-averaged amplitude squared $\overline{|\mathcal{A}|}^2$ (in which the denominator from the propagator contribution is factored out) are, for the scalar and fermion particles, respectively
\bea
\overline{|\mathcal{A}|}_{s}^2&=& 8\left\{(s_1-m_e^2)(s_2-m_e^2)-m_1^2m_2^2)\right\}, \\
\overline{|\mathcal{A}|}_{f}^2&=& 4\left\{ (s_1+s_2)[(m_1+m_2)^2+4m_e^2]\right. \nonumber \\
 &-& (s_1^2+s_2^2)-2m_1m_2(m_1^2+m_2^2+m_1m_2) \nonumber \\
  &-& \left. 2m_e^2(m_1^2+m_2^2+4m_1m_2+3m_e^2)\right\} \, .
\eea
In terms of the variables $s_1$ and $s_2$, defined in Eq.~\eqref{eq:epem}, the total decay width is
\bea
\Gamma_2&=&\frac{g_{12}^2\epsilon^2 \alpha}{64\pi^2 m_2^3 }\int_{s_2^-}^{s_2^+}ds_2\int_{s_1^-}^{s_1^+}ds_1   \\
&\times& \frac{\overline{|\mathcal{A}|}^2}{[m_1^2+m_2^2+2m_e^2-s_1-s_2-m_X^2]^2+m_X^2\Gamma_X^2}\,, \nonumber
\eea
and the integration limits are
\bea
s_1^{\pm}&=&m_1^2+m_e^2+\frac{1}{2s_2}\left[(m_2^2-m_e^2-s_2)(m_1^2-m_e^2+s_2)\right. \nonumber \\
&&\left.\pm \lambda(s_2,m_2^2,m_e^2)\lambda(s_2,m_1^2,m_e^2) \right]\,, \\
s_2^-&=&(m_1+m_e)^2,\hbox{ and }\,s_2^+=(m_2-m_e)^2\,.
\eea

\end{document}